\documentstyle[11pt,newpasp,epsfig, twoside]{article}
\def\edcomment#1{\iffalse\marginpar{\raggedright\sl#1\/}\else\relax\fi}
\reversemarginpar

\begin{document}
\title{TeV $\gamma$-ray Astronomy in the new Millennium}
 \author{Frank Krennrich}
\affil{Iowa State University, Physics \& Astronomy Department
Ames, IA 50011-3160}
  
\begin{abstract}
The field of TeV gamma-ray astronomy is reviewed with emphasis
on its relation to the origin of cosmic rays.  The discovery of TeV
photons from supernova remnants and active galaxies 
 has provided the first direct observational link 
between specific astrophysical objects and particle production 
at the TeV scale. 
TeV $\gamma$-ray observations constrain the high end of the electromagnetic  
spectrum,  a regime most sensitive for testing particle acceleration and 
 emission models.
TeV telescopes have made important contributions to the understanding 
of blazars and supernova remnants, however, it will take the next generation  
atmospheric Cherenkov telescopes and  satellite-based $\gamma$-ray  detectors  
to unravel the mystery of hadronic  cosmic-ray sources.   

A short review of TeV observations is followed by a discussion of the
  capabilities and scientific potential of the next generation 
ground-based atmospheric Cherenkov telescopes.

\end{abstract}

\section{Introduction}

Very High Energy $\gamma$-ray astronomy (VHE defined as the energy 
range of 200 GeV - 50 TeV)  has extended the photon spectrum of
high energy astrophysics of galactic and extragalactic sources
through adding a new observational window.   
Imaging atmospheric Cherenkov telescopes\footnote{CANGAROO, 
CAT, Durham,  HEGRA, Whipple and Telescope Array, for an  overview see 
Catanese \& Weekes (2000)}  
have taken the lead in this energy range because of their large collection area
($\rm 10^{4} - 10^{5}\:  m^2$),  good sensitivity, high 
angular resolution, good energy resolution and low energy 
threshold (Ong 1998). 
The scientific objectives for exploring the universe with VHE  photons  
can be summarized as follows:

\noindent I. One of the long standing prime objectives for VHE $\rm \gamma$-ray astronomy
is to find the sources of cosmic rays.  Cosmic rays can be traced through their 
interactions with matter and subsequent $\gamma$-ray emission via
 $\pi^{0}$ production.  Where and how does nature 
accelerate particles to energies that extend to $\rm 10^{20}$~eV and beyond?

\noindent II. The sky above 10 GeV is still largely unexplored.  This is one of the
last regions of the astrophysical electromagnetic spectrum with the benefit of the
unknown - less than a percent of the sky has
been scanned with highly  sensitive\footnote{A sensitive detector 
is defined in this paper as an instrument capable of detecting the 
Crab Nebula in a few hours of observation at the $\rm >5 \sigma$ level.} 
VHE telescopes.  
It is important to realize that VHE $\gamma$-ray telescopes enter
a regime of particle physics phenomena that is  complementary to accelerator 
laboratories: a)   Annihilation  lines from supersymmetric particle decays
constitute an exciting possibility for astroparticle physics\footnote{Neutralino 
annihilation of astrophysical origin has been recently reviewed  by Bergstr\"om et al. 1998
 as a possible test to constrain supersymmetric particle theory.}. b)   Probing quantum gravity 
effects of photons traveling cosmological distances  
have been suggested (Amelino-Camelia et al. 1998; Biller et al 2000).
c) Also searches for primordial black holes with future imaging atmospheric
Cherenkov telescopes have been revived (Krennrich, Le Bohec \& Weekes 2000).

\noindent III. At lower energies, observations of HE $\gamma$-rays  (20 MeV - 20 GeV)
with the EGRET  telescope on-board of the Compton Gamma Ray Observatory,   have 
revealed more than 270 sources (Hartmann et al. 1999): 
7 pulsars, 66 high-confidence blazars,    27 lower-confidence blazar identifications, 
170  sources not yet identified  with known objects (EGRET unidentified sources).  
It is apparent that the high energy universe  is not as  sparsely populated  as early 
measurements in the  1960s and 1970s with less sensitive first generation  space 
telescopes  OSO-3 (Kraushaar et al. 1972), SAS-2 (Derdeyn et al. 1972; 
Fichtel et al. 1975) and   COS-B (Scarsi et al. 1977) indicated.    
The future generation space telescope  
(Gamma Ray Large Area Space Telescope: GLAST)  is expected to detect in the 
order of $\rm 10^4$ sources (Michelson et al. 1999), with a wealth of  GeV $\gamma$-ray 
astrophysics   to be explored.    A major objective of VHE
$\gamma$-ray astronomy is to extend observations of the electromagnetic 
universe to higher energies,  to arrive at a more complete 
picture of the non-thermal leptonic component in astrophysical objects.

In this paper I emphasize  point I. and  elude to the search for 
evidence for cosmic-ray hadron  acceleration in sources and potential 
sources of VHE $\gamma$-ray emission.     
Recent observations and some selected results relevant to particle acceleration
are summarized in section 2.  
In section 3, I discuss  the next generation atmospheric Cherenkov telescopes, 
their capabilities and their prospects for establishing astrophysical sites 
of particle acceleration, in particular  sources of nucleonic cosmic-rays. 
 
  To date,  imaging atmospheric Cherenkov  telescopes  provide  a source catalog 
(Table 1) consisting of  12 sources (see also Weekes 2000): 
2 high confidence  plerions, 1 lower-confidence   plerion,  2 high confidence blazars, 
 4 lower-confidence blazar identifications,  3  lower-confidence shell-type 
supernova remnants.   
Despite the small number of sources at $\rm E \ge$~200~GeV,  some of the
observations (Crab, Markarian 421 and Markarian 501) have revealed statistically
strong detections\footnote{During flaring activity of Markarian~421 and 
Markarian~501 detections at the 20 - 40 $\sigma$ level were achieved per night
 in a few hours of observation (Gaidos et al. 1996; Quinn et al. 1999).} with  
profound and surprising implications.

\begin{table}
\caption{TeV Source Catalog: Fall 2000}
\begin{tabular}{lrclc} \hline\hline
Source  & Energy  & Flux & Group &   EGRET  \\
        &    (GeV)   & ($\rm \times 10^{-11} cm^{-2} s^{-1} $)  &  & source    \\ \hline
\underline{Plerions} &          &                      &                  \\
 Crab Nebula &   400  & 7.0   &    Whipple, ASGAT, HEGRA  &   pulsed,  \\
          &            &       &       Themistocle, *Gamma,  & unpulsed  \\
                &            &       &      TA, Crimea, CAT  \\
          &            &       &        CANGAROO  \\ 
PSR~1706-44 & 1000   &   0.8   &   CANGAROO, Durham    &  pulsed \\
Vela        &  2500   &  0.29  &  CANGAROO   &   pulsed \\  
          &        &             &                     &    \\    
\underline{Shell-type SNRs} &          &       &                  \\
SN~1006     &  1700   & 0.46   &  CANGAROO   &    no \\
RXJ1713.7-3946 & 1800    &  0.53     &   CANGAROO  &   no  \\  
Cassiopeia A      &   1000     &    -         &   HEGRA  &  no  \\ 
          &        &             &                     &    \\    
\underline{Blazars: XBL} &          &       &                  \\
Markarian 421   & 260   &    variable & Whipple, HEGRA, CAT  & yes  \\
 z = 0.031     &        &             &                     &    \\ 
Markarian 501   & 260   &    variable & Whipple, HEGRA, CAT, TA  & no  \\
 z = 0.034     &        &             &                     &    \\ 
1ES2344+514   & 300   &    variable & Whipple   & no  \\
 z = 0.044     &        &             &                     &    \\ 
PKS2155-304   & 300   &    variable & Durham   & yes  \\
 z = 0.116     &        &             &                     &    \\ 
1ES1959+650   & 600   &    variable & TA   & no  \\
 z = 0.048     &        &             &                     &    \\  
          &        &             &                     &    \\    
\underline{Blazars: RBL} &          &       &                  \\\
3C66A   & 900  &    variable &  Crimea   & yes     \\
 z = 0.44     &        &             &                     &    \\      
\end{tabular}
\end{table}

To answer the question, as to why  so few of the EGRET sources have been seen at TeV
energies, it is useful to  consider galactic and extragalactic sources separately.   
Pulsars and   unidentified galactic EGRET sources  possibly  undergo  an 
intrinsic  cutoff   at the $\gamma$-ray production site  
(e.g., Daugherty \& Harding 1996 and references therein; Grenier 1999).  
For galactic EGRET sources to become detectable with atmospheric Cherenkov 
telescopes and take advantage of their large collection area,  the energy  
threshold of those  has to be lowered, ideally 
down to  the high end of EGRET energies (10~GeV). 
The non-detection  of most EGRET blazars,  might  be related
to an intrinsic cutoff and/or the transparency of the universe on extragalactic
distance scales due to the interaction of TeV photons with the infrared background
photons via $\rm \gamma \gamma  \rightarrow e^{+}e^{-}$.       
 However,  the universe for redshifts of $\rm z \le 1$ becomes essentially
transparent for photon energies E~$\rm \le$~40~GeV (Stecker 2000), showing the need to lower the
energy threshold of atmospheric Cherenkov telescopes 
to the 50~GeV range.

\section{Observations: }

\subsection{Galactic Sources: Supernova Remnants - Plerions}

Plerions constitute  a class of  supernova remnants, 
that   contain a young pulsar with a large spin-down 
rate  that powers a synchrotron nebula 
through the injection of relativistic particles (Harding 1996).
At VHE energies three plerions have been detected:
the Crab Nebula, Vela and PSR1706-44.  For all three sources the emission
is not pulsed and it appears to be constant.  This is different from observations
at GeV energies, where EGRET observed predominantly pulsed emission (Thompson
et al. 1997).   With the GeV observations testing the particle
acceleration process near the pulsar magnetosphere and the VHE observations
constraining  the particle energy distribution and magnetic field in the
surrounding nebula, both are important for the understanding of pulsars.

\subsubsection{\underline{Crab Nebula:}}

 The first unequivocal  detection of VHE photons from the Crab Nebula
(Weekes et al. 1989) has established the feasibility of the ground-based
detection of VHE $\gamma$-rays.      
Production of HE (unpulsed component at E~$\ge 1 $~GeV) and VHE  photons can be explained 
as inverse  Compton scattering by relativistic electrons of target photons from 
synchrotron radiation (synchrotron self-Compton: SSC)  and/or  ambient 
infrared and microwave background photons (inverse Compton: IC). 
Because of its strong synchrotron luminosity (X-rays) and strong magnetic field
(fairly young pulsar),   the VHE $\gamma$-ray  flux from the Crab Nebula is  
attributed dominantly  to  SSC emission (e.g., De Jager \& Harding 1992;
 Harding 1996).  

VHE observations can yield important information about the energy spectrum
of the electrons, thus constraining the particle acceleration mechanism,
e.g., the maximum electron energy.
 VHE  data  (Hillas et al. 1998) together with EGRET data (De Jager et al. 1996) 
and X-ray spectra (see references in Hillas et al. 1998)  have 
been used to estimate the average magnetic field strength in the X-ray nebula.  
A magnetic field of $160 \:  \mu$G seems to provide the best fit to the 
multiwavelength spectrum (Figure 1) between X-ray to several TeV energies (Hillas et al. 1998).
In a  recent paper by Aharonian et al. (2000a)  a magnetic field strength of 
($\rm 170 \pm 30) \: \mu$G has been derived,  confirming the previous result. 
The synchrotron self-Compton model also allows the derivation of the
maximum electron energy (De Jager \& Harding 1992) of $\rm E_{max} \sim  10^{16}$~eV.

\begin{figure}
\plotone{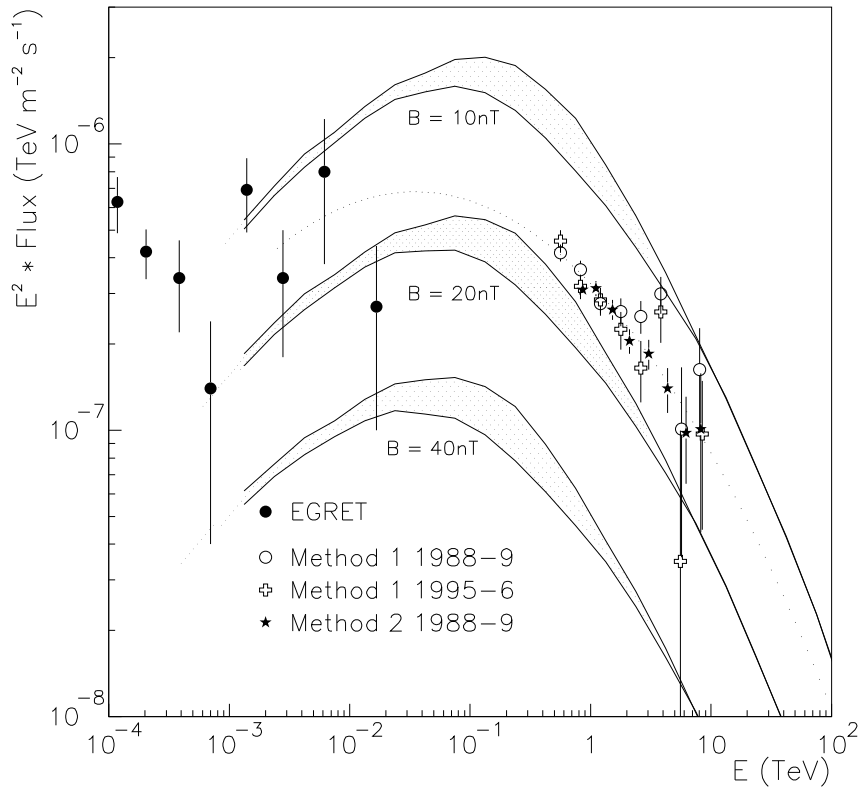}
\caption{Whipple (Hillas et al. 1998) and EGRET (Nolan et al. 1993) measurements of 
the Crab Nebula $\gamma$-ray spectrum
in the HE - VHE regime are shown.  The dotted line is a fit to the Whipple and EGRET fluxes 
using an SSC-model fitting the X-ray and $\gamma$-ray spectrum consistently under the assumption of
an average  magnetic field of $\rm 160 \: \mu$G in the X-ray nebula (Hillas et al. 1998).  }
\end{figure}
No evidence   for $\gamma$-rays from $\pi^{0}$ decays is present in the data.
If protons were accelerated in the Crab Pulsar (Atoyan \& Aharonian 1996;
Bednarek \& Protheroe 1997),   they could 
make a noticeable contribution to the $\gamma$-ray flux above 10~TeV, a regime where
 synchrotron cooling and the energy dependency of the Klein-Nishina 
cross-section  steepens the SSC  contribution.  

The measurement of the Crab spectrum above 10 TeV can be most efficiently 
addressed using the large zenith angle technique  (Krennrich et al. 1997; 
Krennrich et al. 1999a; Tanimori et al. 1999; Aharonian et al. 2000a).   A dedicated program 
over the typical lifetime (5 years) of one of the future telescope 
arrays (VERITAS) could achieve a high precision spectrum of the 
Crab Nebula between 50~GeV and  50~TeV. 

At the low energy end, a recent detection of 50~GeV photons from the Crab 
Nebula by the CELESTE  collaboration (De Naurois et al. 2000) using 
an array of heliostats as a  $\gamma$-ray telescope emphasizes the 
progress that has been made in closing the energy gap between satellite instruments 
and ground-based telescopes.
The Crab Nebula will play an important role in cross-calibrating future 
 ground-based $\gamma$-ray telescopes located in the  northern hemisphere  
(MAGIC, VERITAS) with GLAST. 

In summary, the Crab Nebula although not favored as an acceleration site
of hadronic cosmic rays, is an important testbed and calibration source
for models (in particular SSC) of $\gamma$-ray production and 
for searching for a hadronic component at $\rm E >$~10~TeV.

\subsubsection{\underline{Other plerions: Vela, PSR 1706-44}}

Vela (Yoshikoshi et al. 1997) and PSR 1706-44 (Kifune et al. 1995: Chadwick et al. 1997)
have been detected in VHE $\gamma$-rays. 
The VHE $\gamma$-rays from the direction of $\gamma$-ray pulsar PSR 1706-44 could be 
 associated with a surrounding nebula similar to the Crab Nebula.  However, the 
emission is likely due to inverse Compton scattering of electrons off the 2.7~K 
cosmic microwave background radiation.  Since unpulsed X-ray emission has been observed 
by ROSAT (Becker et al. 1994) a plerion with a weak magnetic field seems a 
reasonable explanation for the VHE emission from PSR 1706-44. 
Plerions with a weak magnetic field could become standard candles in the HE - VHE
 $\gamma$-ray regime to test  models involving IC scattering on the cosmic microwave 
background.

\subsection{Galactic Sources: Pulsars}
 
A major distinction between pulsar models is whether the pulsed $\rm \gamma$-ray emission 
 is produced by particles accelerated near the polar cap
(Daugherty \& Harding 1996) or in the outer magnetosphere 
(Cheng, Ho \& Ruderman 1986; Romani 1996).
Both mechanisms result in $\rm e^{+}e^{-}$ pairs, which produce synchrotron 
radiation and,  via  Compton scattering, boost   soft photons to high energy 
$\rm \gamma$-rays.   If VHE photons were produced  near the polar cap, 
they  would produce $\rm e^{+}e^{-}$  on the strong magnetic fields, 
precluding the detection of VHE $\gamma$-rays. 
Therefore,  if TeV pulsations are observed, then pulsed emission must have its
origin   relatively   far from the neutron star surface, e.g., in the
outer magnetosphere, supporting the outer gap model.

\subsubsection{\underline{Crab, Geminga, Vela, PSR B1951+32:}}
EGRET has detected at least 6 pulsars: Crab, Geminga, Vela,  PSR B1951+32,
PSR1706-44, PSR B1055-52 (Thompson 1997).
Searches for pulsed emission at VHE energies from the Crab Nebula
 (Lessard et al. 2000; Aharonian et al. 1999a), Geminga (Akerlof et al. 1993;
Aharonian et al. 1999a), the Vela pulsar  (Yoshikoshi et al. 1997),
PSR1509-58 (Sako et al. 2000) and  PSR B1951+32 (Srinivasan et al. 1997) 
have yielded null results.  
EGRET data of PSR B1951+32 in fact shows a rising spectrum  indicating that 
the maximum power occurs at several GeV.  The Whipple upper limit 
(Srinivasan et al. 1997) to the pulsed  flux (E$\rm > 260$~GeV) is two 
orders of magnitude below the extrapolated   EGRET spectrum setting the
most severe constraint on the outer gap model (Cheng, Ho \& Ruderman 1986).
Instruments with a lower energy threshold ($\rm <$~50 GeV) are required  
(De Jager 2000) to constrain the  apparently sharp cutoff in pulsars
through a measurement of the pulsed $\gamma$-ray spectrum.

\subsection{Galactic Sources: Shell-type Supernova Remnants }

Supernovae are  the primary candidates held responsible for the flux of hadronic 
cosmic rays  up to energies of approximately $\rm  Z \times 10^{14} eV$ 
(Z = nuclear charge or particle) for two reasons:    1. they appear to be 
the only galactic  objects capable of supplying the power required for 
the cosmic-ray energy density in our galaxy, 2.  a theory of diffuse 
shock acceleration  (Blandford \& Ostriker 1978; Bell 1978; Legage \& Cesarski 1983) 
does produce a  power-law  spectrum of $\rm dN/dE \propto E^{-2.1}$.   
This would be consistent with the observed local cosmic ray spectrum  
$\rm dN/dE \propto E^{-2.7}$,    after correcting for galactic diffusion  
by  $\rm  \propto E^{-0.6}$  (Swordy et al. 1990). 
Biermann \& Strom (1993) suggest a primary source spectrum somewhat steeper 
($\rm dN/dE \propto E^{-2.33}$)  requiring  a factor of $\rm  
 dN/dE \propto E^{-0.3}$ for galactic leakage.

It was suggested (Drury, Aharonian \& V\"olk 1994;  Naito \& Takahara 1994) 
that observations of HE - VHE $\gamma$-rays,  due to  collisions of cosmic ray nuclei
with the interstellar medium  via $\pi^{0}$ production,  could provide the 
crucial evidence for cosmic-ray acceleration  in shell-type supernova remnants (SNRs). 
The  $\gamma$-ray spectrum would  reflect the spectral  index of the 
cosmic-ray spectrum at the source. 
A clear indication for acceleration of nuclei in SNRs
would be  the $\rm \pi^{0}$  bump at low energies with the spectrum extending to 
ten's of TeV, assuming the background from the galactic plane diffuse 
$\gamma$-ray emission can be separated from the SNR. 

However, experimentally the situation is complicated by a possible inverse Compton
$\gamma$-ray component from VHE electrons boosting 2.7~K microwave background photons 
to TeV energies. 
In order to establish $\gamma$-rays with $\pi^{0}$ origin in supernova remnants, the 
contribution from electrons has to be subtracted.   This will require to measure the 
$\gamma$-ray energy spectrum over a wide energy range, ideally from 
10~MeV - 10 TeV.

\subsubsection{\underline{SN 1006, RXJ1713.7-3946:} }

Recently the CANGAROO collaboration has reported the detection of $\gamma$-rays from 
two  shell-type SNRs - SN 1006 (Tanimori et al. 1998b)\footnote{It should be noted that
that observations (Chadwick et al. 1999) with the Durham Mark VI telescope set 
an upper limit on this source conflicting
 with the CANGAROO observation. For further details see Weekes (1999).}, 
and RXJ1713.7-3946 (Muraishi et al. 2000).  
  These observations still require  confirmation and the measurement
of  energy spectra, accurate enough to distinguish a spectral index (2.1) 
for  diffuse shock acceleration (Blandford \& Ostriker 1978; Bell 1978; 
Legage \& Cesarski 1983) from other scenarios. 

  To date, these detections prove to be  inconclusive as to whether or  not 
shell-type SNRs accelerate hadronic cosmic rays  (Aharonian \& Atoyan 1999; Baring 1999). 
The reason is, that the origin of the TeV photons can be linked to the X-ray  
emission in these objects.    The discovery of non-thermal 
X-ray emission  in SN 1006 provided the first unambiguous evidence for electrons
with energies up to 100~TeV in SN~1006 (Koyama et al. 1995). 
Similarly SNR RXJ1713.7-3946 was recently discovered as an X-ray source
in  ROSAT and ASCA data (Pfeffermann \& Aschenbach 1996). 

Therefore,  VHE photons  detected  from  SN 1006 and RXJ1713.7-3946  may well arise
from  VHE electrons via inverse Compton scattering against low energy photons (2.7~K).
In fact, VHE emission from SN 1006 had been suggested by several theorists (Pohl 1996;
Mastichiadis 1996; Mastichiadis \& de Jager 1996; Yoshida \& Yanagita 1997) based on 
this picture using the X-ray luminosity and an equipartition magnetic field estimate. 
Conversely, the VHE flux together with  the X-ray flux from SN 1006 
provides an estimate of the magnetic field strength in the X-ray shell, 
yielding a value of $\rm 6.5 \pm 2 \mu$G (Tanimori et al. 1998b).   
A hadronic component is however not ruled out, but might  be difficult to
distinguish from the IC component based on the VHE spectrum alone (Aharonian \& Atoyan 1999). 
These authors suggest that spatially resolving the VHE emission region to search for
correlations with matter density might become an important tool in settling
 the question whether or not hadronic processes  play a significant role. 

Another example of a shell-type supernova remnant with a potentially strong inverse Compton component
is RXJ1713.7-3946 because it has been detected  in X-rays (Pfeffermann \& Aschenbach 1996) 
showing a non-thermal emission in the shell. 
A detection of  VHE photons has been reported by  Muraishi et al. (2000). 
 Under the assumption that the  origin of the TeV photons is related to  IC scattering of electrons 
on the microwave background radiation,  a magnetic field of $\rm 11 \: \mu$G was estimated.\


\subsubsection{\underline{IC443,  $\gamma$-Cygni,  W44, W51, W63 and Tycho:} }

Neither SN 1006 nor RXJ1713.7-3946 were on the top of the list of good candidates for
searching for VHE emission from neutral pions  in  shell-type SNRs.  This is different for
a set of SNRs chosen by Buckley et al. (1998):  IC443,  $\gamma$-Cygni,  W44, W51, W63 and Tycho.
Among other criteria, the latter were selected with view for an enhanced $\pi^{0}$ component through
 a possible association with a molecular cloud.  Most importantly,  $\gamma$-Cygni, W44 \& IC443 
 show a possible association with  EGRET sources: the positions of 3 unidentified EGRET sources  
are consistent with  these shell-type SNRs  
(Sturner \& Dermer 1995; Esposito et al. 1996; Lamb \& Macomb 1997; 
 Jaffe et al. 1997).  However, because of EGRET's limited angular resolution a 
clear identification  cannot be made.
 
Following Buckley et al. (1998), by assuming that the EGRET emission is from these shell-type 
SNRs and that $\gamma$-ray emission  comes dominantly from $\pi^{0}$'s,     the 
spectral  index for these SNRs would have to be softer
(2.5 for $\gamma$-Cygni, 2.4 for IC443) than predicted by  Drury, Aharonian \& V\"olk (1994). 
 Gaisser, Protheroe \& Stanev (1998) performed multiwavelength fits to the EGRET data and Whipple upper
limits.   They find that a differential spectral index of nuclei and electrons (assuming 
electrons and nuclei are accelerated with the same spectral indices) is $\rm \sim$ 2.2 - 2.4
(Gaisser, Protheroe \& Stanev 1998).

\subsubsection{\underline{Cas-A:} } 

A high  X-ray flux and  a hard spectrum (Allen, Gotthelf \& Petre 1999) and the relatively 
high ambient matter density makes Cas-A an example of a shell-type supernova remnant that 
might exhibit both,  a strong  IC and/or a high $\pi^{0}$ $\gamma$-ray  component. 
HEGRA presented evidence for VHE emission from Cas-A (P\"uhlhofer et al. 1999) at the 
3.4 - 4.5 $\sigma$ level, however,  verification of the result will be required.
  
GLAST and future ground-based Cherenkov telescopes will provide unambiguous source 
identification and measurements of the energy spectra with high sensitivity.  GLAST at the low
end of the spectrum could see the $\pi^{0}$ bump (Ormes, Digel, Moskalenko \& Williamsen 1999) which could
provide evidence for acceleration of hadronic nuclei.  However,  the background from the 
diffuse galactic plane emission also exhibits a $\pi^{0}$ bump,  making this signature
ambiguous.   TeV telescopes could measure the  spectrum between 50~GeV 
up to ten's of TeV providing together with GLAST a spectral constraint between
10 MeV - 50~TeV.   Together with the X-ray spectrum this could provide the spectral
data necessary to untangle the IC from the nucleonic components.   
Furthermore spatial correlations with X-ray emission could trace the IC component whereas
a correlation with matter density could trace the nucleonic component.

\subsection{Galactic Plane: }

Cosmic rays interacting with the interstellar medium (ISM) give rise to the
emission of $\gamma$-rays over a wide range of energies. The  processes considered
 are bremsstrahlung, inverse Compton scattering, and the $\pi^{0}$
production via hadronic interactions of protons and nuclei; however, the magnitude
of their individual contributions to the observed $\gamma$-ray flux has become
a matter of debate.
In  the COS-B era, the situation appeared fairly non-controversial
with the $\gamma$-ray emission (Mayer-Hasselwander et al. 1982) 
matching the model by Bertsch et al. (1993) well without major 
discrepancies between observations and theory.

This has changed with EGRET  observations.  Although,  the EGRET 
data show good spatial agreement with  model calculations, at energies 
above 1~GeV the observed intensity surprisingly exceeds the model prediction
by 60\% (Hunter et al. 1997).   
Various suggestions have been made to resolve this discrepancy.  Hunter et al. (1997)
proposed  that the excess is due to unresolved point sources.  Pohl et al. (1997) showed that
unresolved $\gamma$-ray pulsars  could significantly contribute to the
diffuse galactic emission; however,  the latitude distribution for pulsars
is too narrow to explain the observed excess.

\begin{figure}
\plotone{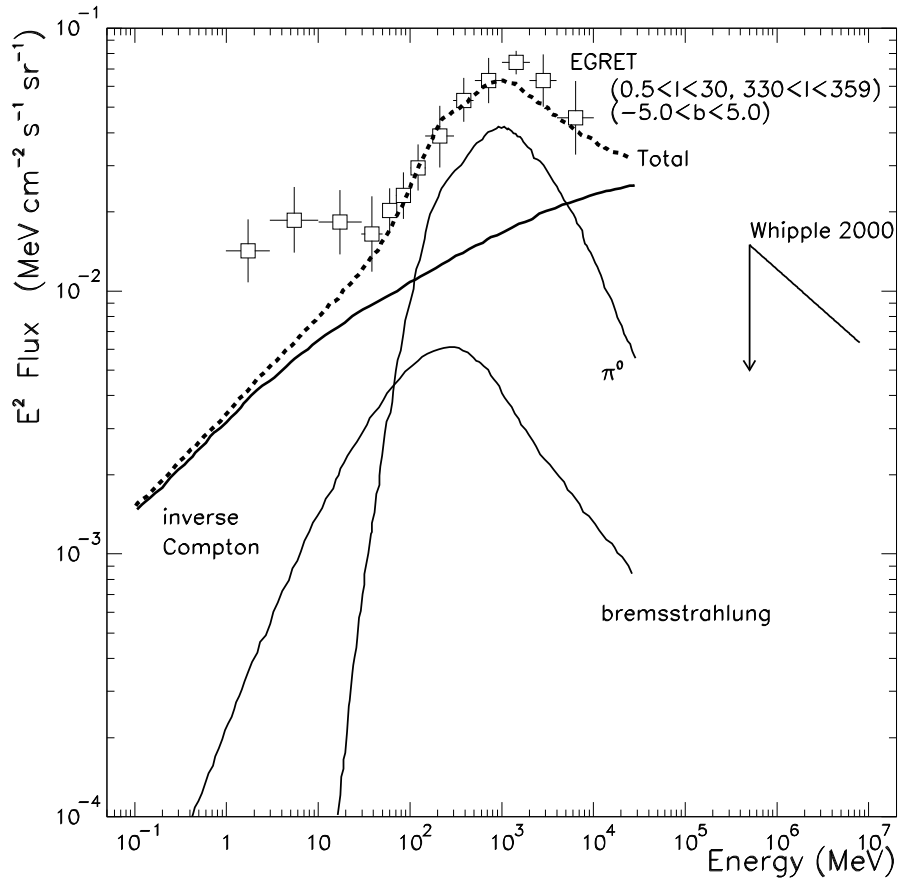}
\caption{The energy spectrum of the diffuse galactic plane emission as measured by EGRET
(Hunter et al. 1997) is compared to a  model (HEMN model) by Strong, Moskalenko \& Reimer (2000). 
This model is based on an electron injection index of $\rm \alpha = 1.8$ ($\rm dN/dE \propto E^{-\alpha}
$) and a modified nucleon spectrum.
Furthermore, we show the upper limit from Whipple observations (Whipple 2000) of the galactic plane
at $l = 40^{\circ}$, $-2.0^{\circ} < b <  2.0^{\circ}$  by Le Bohec et al. (2000).}
\end{figure}

Attempts to explain the flux at  E~$>$~1~GeV  by a harder interstellar proton spectrum
(Mori 1997; Moskalenko, Strong \& Reimer 1998)  than locally observed, are limited by 
constraints from  antiproton and positron  measurements (Strong, Moskalenko \& Reimer 2000).
It has also been suggested (Pohl \& Esposito 1998), that the interstellar electron spectrum 
can be harder than that locally observed, allowing a significant inverse Compton 
contribution at higher energies.     A recent  study  
(Strong, Moskalenko \& Reimer 2000) shows in fact that a harder electron  spectrum 
(injection index 1.8) and a modified nucleon spectrum (in agreement with antiproton and
positron flux constraints) can fit the data
between 10~MeV - 30~GeV reasonably well (see Figure 2).
Given the hard $\gamma$-ray spectrum above 1~GeV  the prospects for 
detecting the diffuse emission  from the galactic plane at 100's of GeV with ground-based 
Cherenkov telescopes are promising if the spectrum extends.  Observations by the 
Whipple collaboration  (Le Bohec et al. 2000) have provided an upper limit that is 
close to the extrapolation  of the EGRET spectrum.  These data resulted in a 
lower limit of $\rm \alpha = 2.31$ ($\rm dN/dE \propto E^{-\alpha}$) on the differential  
spectral index (Figure 2),  assuming there is no break 
in the spectrum between 30 - 500~GeV. 
The  Whipple upper limit at 500~GeV, provides strong  evidence that,  if the excess at GeV
energies is due to an IC component from electrons, either the electron spectrum undergoes
a sharp intrinsic cutoff  at the acceleration site, or cooling processes result in a 
spectral break.

\subsection{Extragalactic Sources: Blazars}

Active galactic nuclei (AGNs) observed at $\gamma$-ray energies of
E $\rm > $ 100~MeV are believed to exhibit highly anisotropic
radiation along their jets.  Those with their jet axis closely aligned
with the observer's line of sight  are collectively called blazars
and include optically violent variable quasars, highly polarized
quasars and BL Lacertae (BL Lac) objects.  Their emission spectrum
is dominated by non-thermal emission which spans the entire wavelength 
range from radio to $\gamma$-rays. Short  flux  variability 
and dominantly non-thermal emission suggests that the observed radiation in these 
objects is produced primarily by a jet of highly relativistic  particles with the
emission region moving at relativistic speed towards the direction of the observer.

This preferred geometry implies that the luminosity in the direction of the jet is 
greatly enhanced by the 4th power of  the relativistic Doppler factor (typically $\rm \sim \: 10$),
 making blazars appear even more  powerful  (Blandford \& Rees 1978).

\subsubsection{\underline{Variability: Markarian~421, Markarian~501:} }

VHE observations of Markarian 421 and Markarian 501 have revealed extremely variable
$\gamma$-ray  emission (Gaidos et al. 1996: Quinn et al. 1999).   Observations
of Markarian 421 on May 7 1996  came as a surprise,  showing a flux increase by a 
factor of 50 reaching a  maximum of 10 Crab (flux measured in units of  the steady 
emission from the Crab Nebula) within  two hours (Figure 3).  
A remarkably short doubling and decay time of  15 minutes  was observed on May 15 1996.  
These short flux variations provide information about particle acceleration, energy loss,
the emission process in the jet, the bulk Lorentz factor (from $\gamma\gamma \rightarrow e^{+}e^{-}$ 
opacity argument)   and the size of the emission region.
The Markarian 421 observations indicate a compact emission region of 1 - 10 light
hours in diameter ($\rm R \le 10^{-4}$ pc) corresponding to 10 Schwarzschild radii 
of  a $\rm 10^{8}$ solar mass  black hole (Gaidos et al. 1996).

\begin{figure}
\plotone{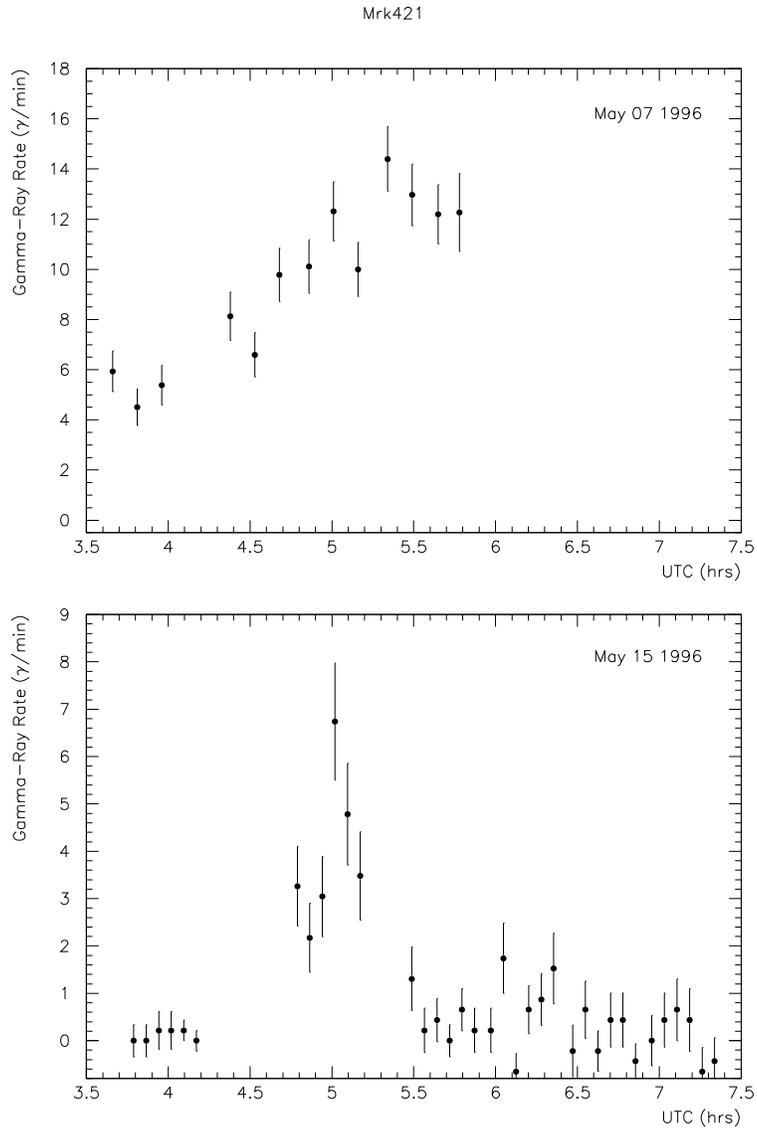}
\caption{VHE $\gamma$-ray lightcurves  of two rapid flares from Markarian 421 on May 7th and May 15th
 1996 (Gaidos et al. 1996). }
\end{figure}

The  location of the emission region could be either close to the supermassive black 
hole or, in the case of a shock front propagating out along the jet, substantially further
out.   It is important to note, that the minimum detectable variability time scale
is limited by the sensitivity of existing $\gamma$-ray detectors, shorter variability time
scales cannot be ruled out.   One objective of future generation detectors will be to measure the
minimum variability time scale, by the means of a  lowered energy threshold and increased
collection area.  

Variability over a wide range of  time scales has been observed for the $\gamma$-ray emission
of Markarian 501.  Markarian 501 at the time it was  discovered as a $\gamma$-ray blazar
(Quinn et al. 1996) showed a flux level of $\rm \sim$  7\% of the  Crab.   
Figure 4 shows the night by night $\gamma$-ray rates between 1995 - 1998 in units of Crab.  
A variability time scale of a month can be seen in the 1996 data,
 day-scale  variability (up to $\rm \sim$ 4 Crab) can be seen in the 
1997 data and even hour scale variability can be seen for two nights in 1997 
(Quinn et al. 1999).  

The wide range of variability time scales established by these observations naturally 
raises the question as to what causes these variations and further more, are they of
the same origin?
Perhaps the most intriguing questions about the particle jet are, what is the
acceleration mechanism and which particles are accelerated in the jet: electrons
and positrons (collectively called electrons), or protons, or both.
In leptonic models of blazars,    the main source of $\gamma$-rays is Compton scattering 
of soft photons by  energetic electrons in the jet.   The source of soft photons 
could be due to self-synchrotron radiation of the electrons in the jet 
(Maraschi, Ghisellini \& Celotti  1992; Marscher \& Bloom 1994),   referred to 
as SSC models.   On the other hand, if the soft photons originate  from  ambient 
radiation external to the jet,  e.g., the accretion disk (Dermer, Schlickeiser \& Mastichiadis 1992) 
or broad-line region clouds  (Sikora, Begelmann \& Rees 1994; Blandford \& Levinson 1995).
Modelers call  those external Compton models (EC).

A  source of $\gamma$-rays related to cosmic ray acceleration could be due 
to photo-meson  ($\rm p \gamma \rightarrow \pi X$) or photo-pair 
($\rm p \gamma \rightarrow e^{\pm} X$)   production of high energy protons 
 (Mannheim \& Biermann 1992).  The photo-meson and photo-pair 
production cross sections are a factor of $\rm \sim  10^3$  smaller than the 
typical cross section of the leptonic processes requiring  higher 
proton energies (Lorentz factors) in the range
of $\rm 10^{17} - 10^{19}$~eV (Mannheim 1993) to explain VHE emission from blazars.
Also synchrotron radiation from protons has been suggested as a source of TeV photons
(M\"ucke \& Protheroe  2000; Aharonian 2000).

Strong flux variability can be used to estimate the cooling time scales and/or the 
acceleration time scale  (whichever dominates the process) of the
emission mechanism at work,   providing an additional constraint to models (Maraschi et al. 1999). 
For example, the high energy radiation from a photo-meson or photo-pair
induced cascade decays more slowly than  SSC radiation  (B\"ottcher \& Dermer 1998).
However, testing these models requires that the spectrum between  X-rays  and the 
TeV emission  is measured on short time scales, as short as the variability time scale
 to better constrain the mechanism.    This requires contemporaneous 
observations  over a wide range of energies - X-rays, HE  and VHE $\gamma$-rays - 
the  so-called multiwavelength campaigns.

\begin{figure}
\plotone{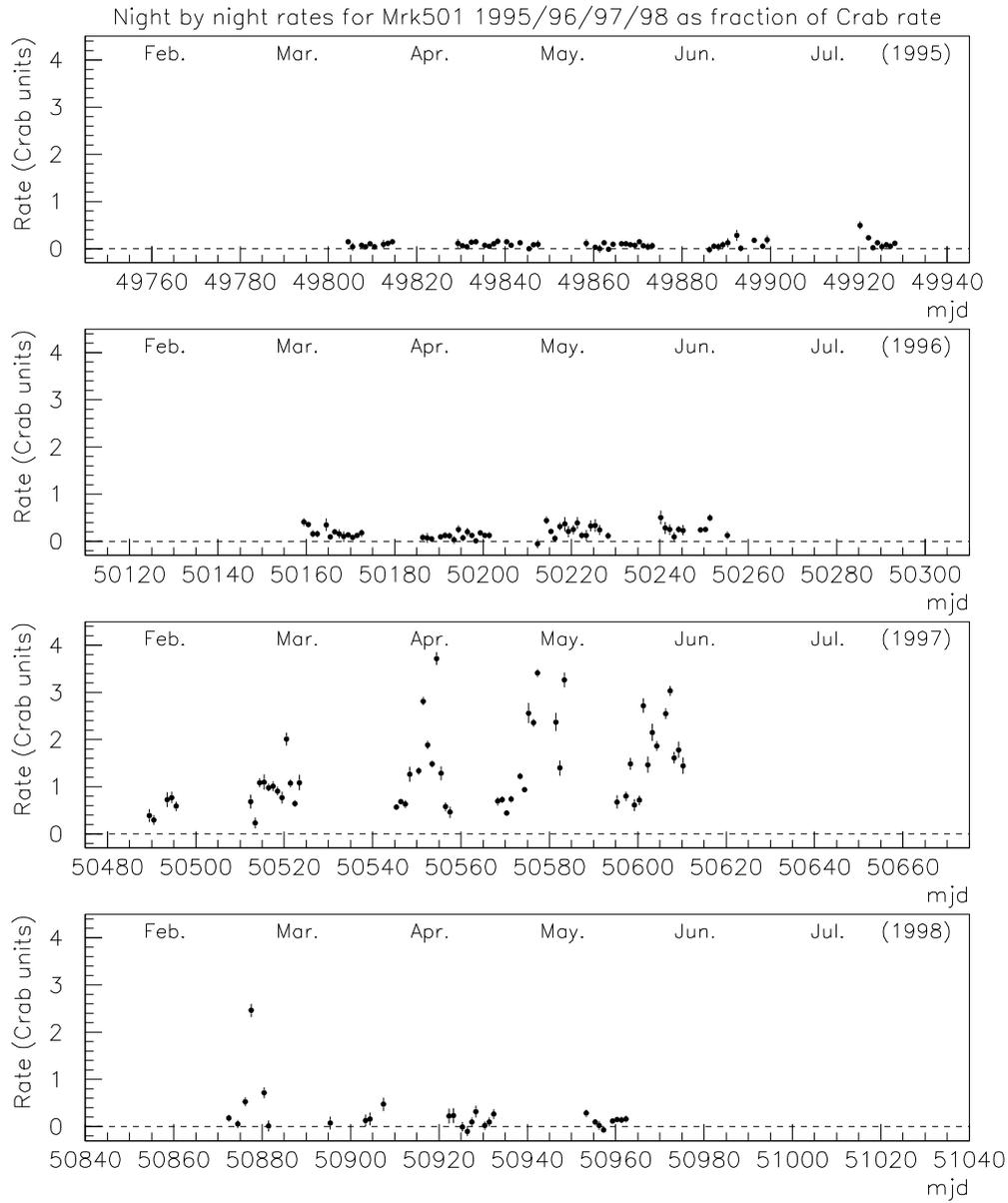}
\caption{The nightly fluxes for Markarian 501 observed with the Whipple telescope 
are presented for the years 1995 - 1998.  Figure is from Quinn et al. (1999). }
\end{figure}

\subsubsection{\underline{Multiwavelength Observations:} }

After the first indication for correlation of a TeV flare with an X-ray flare in 
1995 (Macomb et al. 1995; Macomb et al. 1996) a series of multiwavelength 
campaigns involving the Whipple telescope   and X-ray telescopes were initiated.   
In a campaign on Markarian 421, Whipple/ASCA observations  between 
April 20 - May 5 1995  showed convincing evidence for correlations between TeV
and X-ray photons (Buckley et al. 1996).    
These first correlated TeV/X-ray  observations were undersampled and, therefore,  not strictly 
contemporaneous.    Given the short variability time scales,  
simultaneous observations are crucial  to establish X-ray/TeV correlation. 
Thus, a concerted effort was undertaken by VHE telescope groups (CAT, HEGRA, and Whipple) 
to provide better sampling of observations.   

A campaign in late April 1998 by Whipple/BeppoSAX  revealed a short flare 
(flux halving time of 0.94 hours) at 2 TeV with a   contemporaneous  X-ray flare 
(flux halving time of 20.4 hours).    
The two flares are simultaneous to within 1.5 hours.   These observations 
(Maraschi et al. 1999) provide the first evidence that the X-ray and TeV
emission is well correlated on time scales of hours (Figure 5). 
Simultaneity of the two flares implies that the X-ray and the TeV photons arise from
the same emission region,  likely from the same population of synchrotron 
radiating electrons.  This observation supports the notion that the SSC
mechanism is at least partially if not dominantly  at work in the  $\gamma$-ray
production in  Markarian 421.
   In addition it was possible, for the first time to derive detailed energy spectra for 
a contemporaneous X-ray and TeV flare (Figure 6).   By assuming an SSC model the spectra 
can be fitted yielding an accurate estimate of the physical parameters  
(magnetic field strength B = 0.06 G, radius of emission region R = $\rm 10^{16} \: cm$, 
Doppler factor $\rm \delta$ = 20).     However, the HE-VHE part of the 
spectrum needs a more sensitive and wider energy coverage\footnote{An
energy range of  GeV - ten's of TeV would be desirable.}  to tightly 
constrain  the mechanism and rule out some of the models.  

An X-ray/TeV campaign involving the Whipple, CAT and HEGRA telescopes 
and ASCA at X-rays shows for the first time a series of flares
continuously covered (for 1 week) at X-rays (Figure 7) and good sampling at VHE energies 
(Takahashi et al. 1999).   Again it is  apparent that a better sensitivity with 
TeV telescopes would be required to test emission models in detail.  However,
observations like these indicate the good prospects for studying X-ray/TeV
correlations on time scales of minutes - hours - days with multi-telescope
installations located at different geographical longitude and continuous X-ray 
sampling to fully constrain the emission  process.

Blazar multiwavelength studies including  VHE $\gamma$-rays would be of limited interest if they
 would  involve just  a single source with little implications on blazars in general.
  The discovery of Markarian 501 as a  strong  $\gamma$-ray emitter peaking at
TeV energies has promoted  TeV blazar observations beyond the status of a
 fringe science.   
Flaring activity of Markarian 501 started in February 1997 (Protheroe, Bhat, Fleury, Lorenz 1997) 
lasting until October 1997 (see Catanese \& Weekes 2000), inspiring numerous
observations including X-ray, MeV - GeV  and various TeV telescopes.
Figure 8 shows the result of a multiwavelength campaign (Catanese et al. 1997;
Pian et al. 1998) and the following properties are significant as to the role
of TeV blazar observations:

\begin{figure}
\plotone{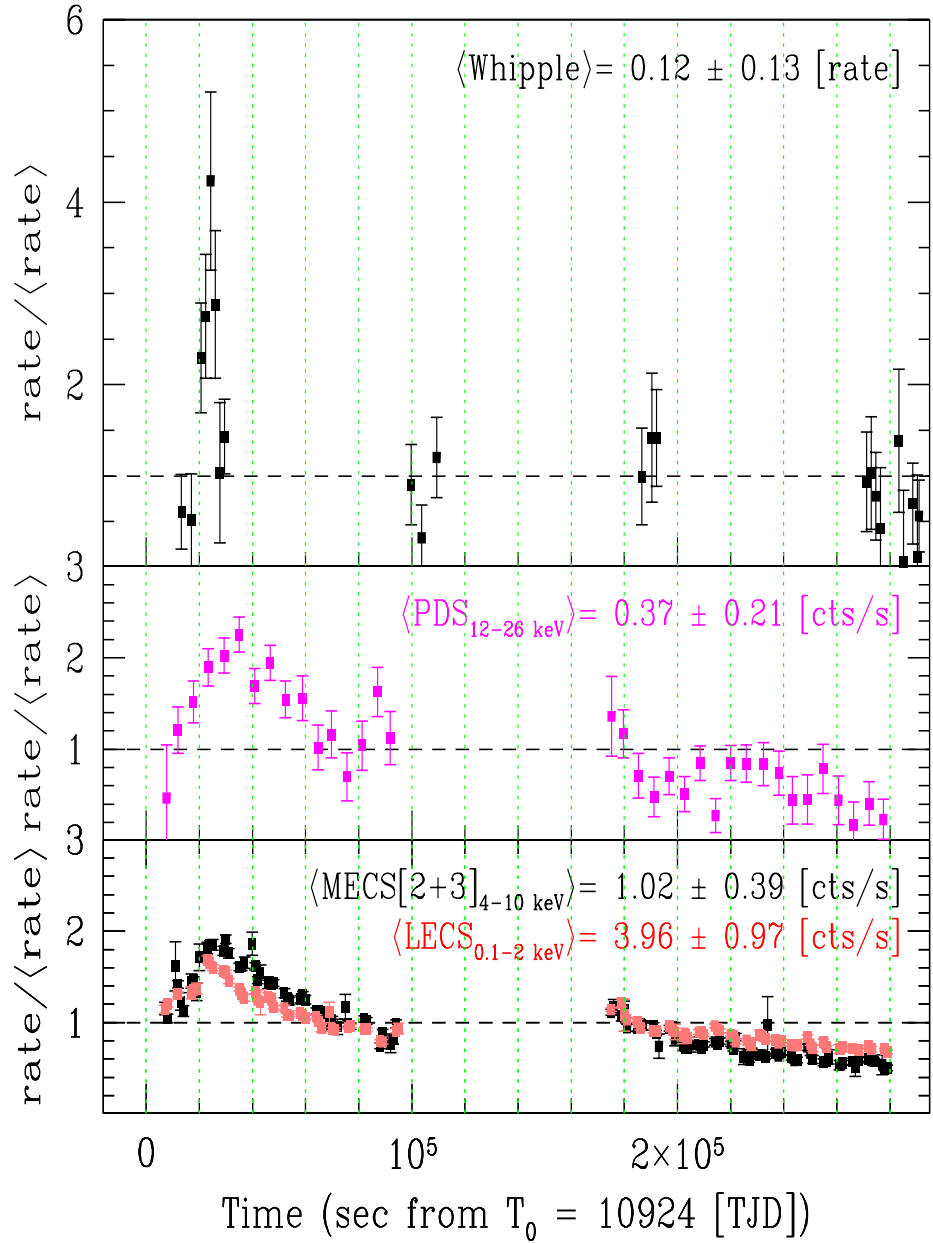}
\caption{The lightcurve of Markarian 421 during a short flare on  April 21,  detected by Whipple 
and BeppoSAX 21 from  Maraschi et al. (1999).  The top lightcurve shows the Whipple $\gamma$-ray rate
normalized to the average rate over the observations shown.   The lightcurves below  
are for various energy ranges of the BeppoSAX data.   }
\end{figure}

\begin{figure}
\plotone{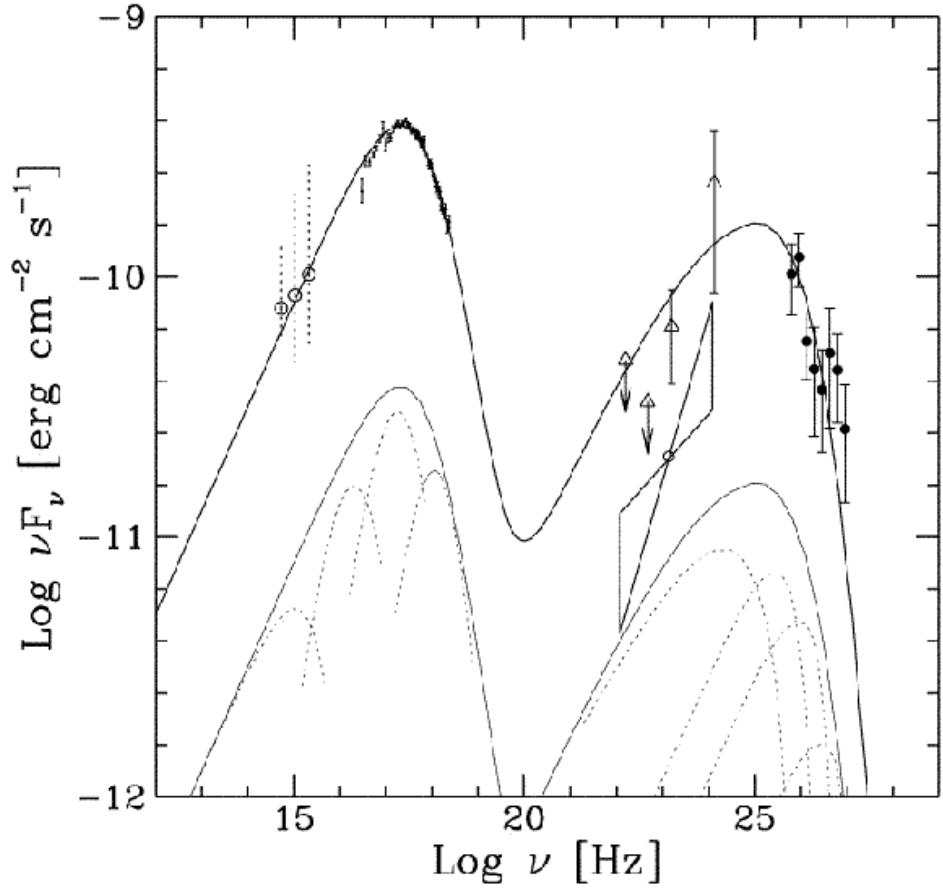}
\caption{The multiwavelength spectrum of Markarian 421 showing contemporaneous 
spectral data from a short flare  in the X-ray and VHE regime (Maraschi et al. 1999).  
The solid line is a fit to the spectrum based on the SSC model.   The Figure is from
Maraschi et al. (1999). }
\end{figure}

\begin{figure}
\plotone{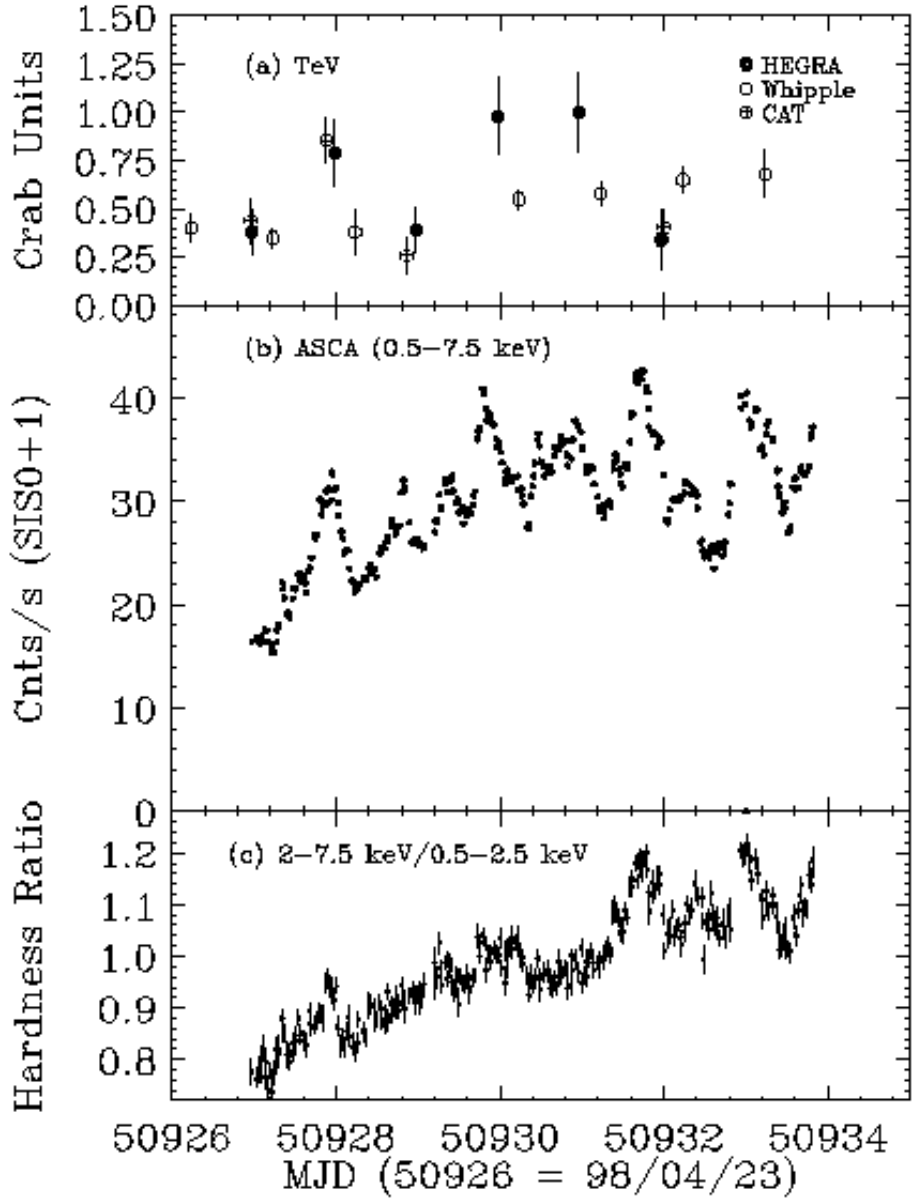}
\caption{Observations of Markarian 421 taken in 1998 April - May:  a) TeV data
from CAT, HEGRA and Whipple,  b) X-ray flux observed with ASCA,  c) X-ray hardness ratios 
observed with ASCA are shown.  Figure is from Takahashi et al. (1999). }
\end{figure}

\begin{figure}
\plotone{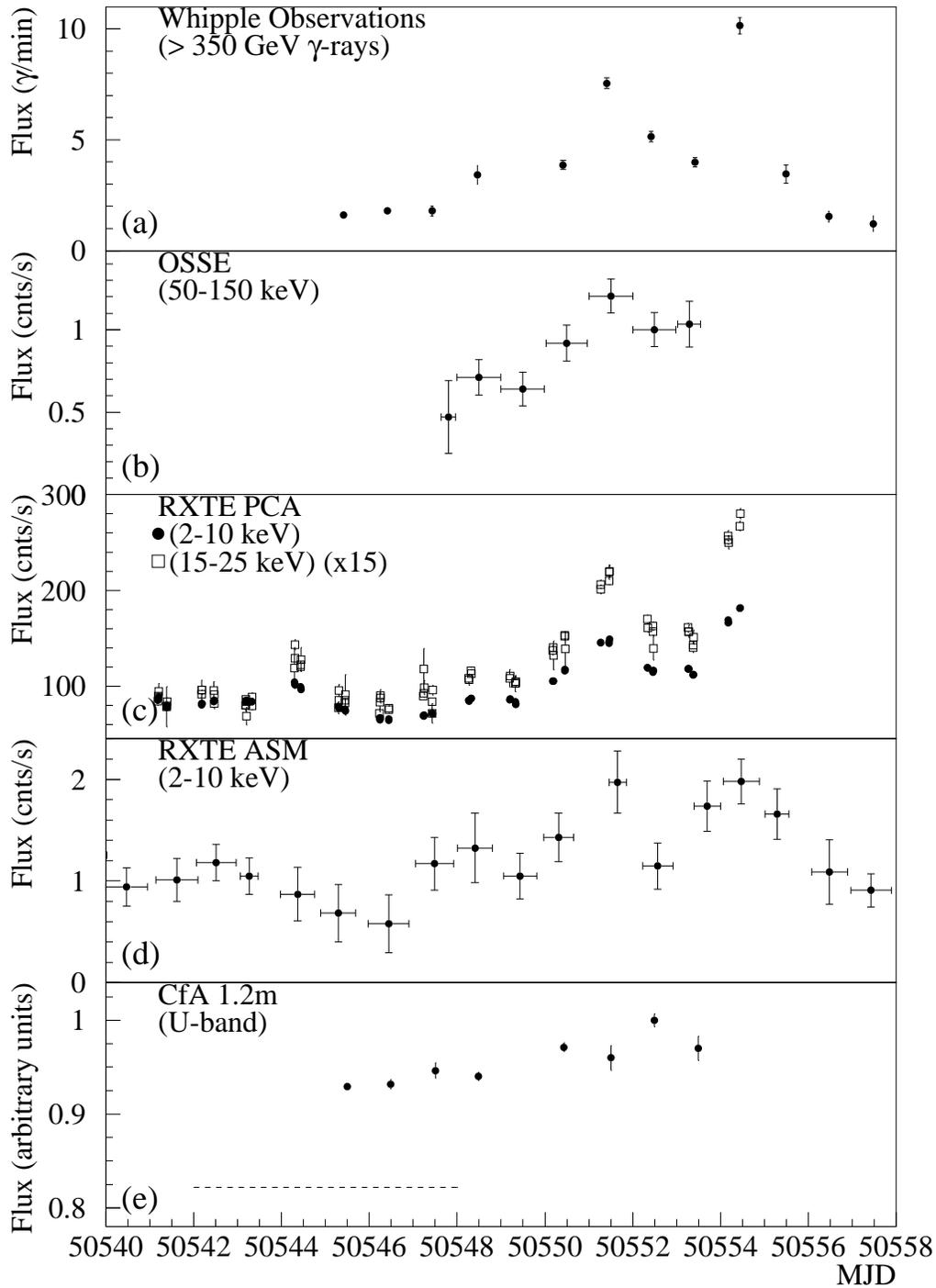}
\caption{Multiwavelength observations of Markarian 501 (Catanese et al. 1997) taken during 
strong flaring activity in April 1997 are shown:  a) VHE $\gamma$-ray, b) hard X-ray, c) soft X-ray,
d) U-band optical.  }
\end{figure}

\begin{figure}
\plotone{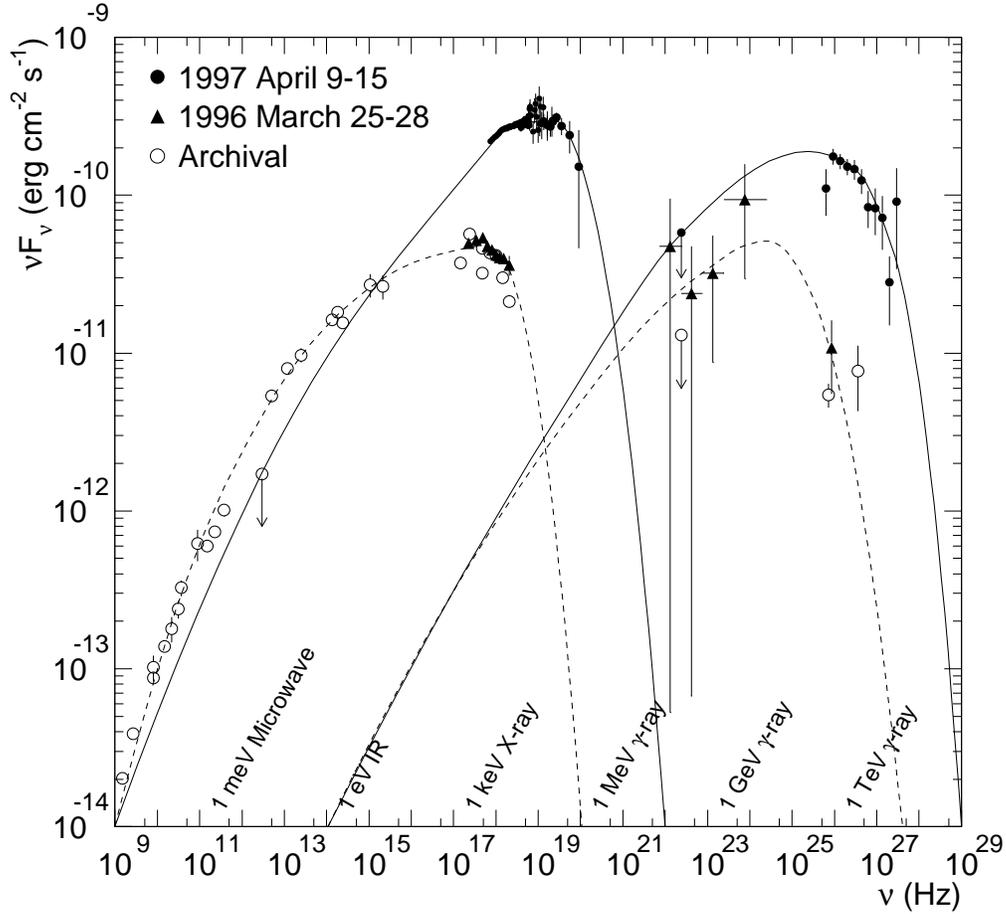}
\caption{Multiwavelength spectrum of Markarian 501 (Catanese \& Weekes 2000) from
contemporaneous and archival data (Catanese et al. 1997; Catanese (1999); Kataoka et al. (1999)).
The Figure is from Catanese \& Weekes (2000).  }
\end{figure}

 1) The TeV flaring activity coincides with a strong detection by the OSSE
detector aboard the CGRO between 50 - 470 keV.  OSSE has detected only a few blazars
(McNaron-Brown et al. 1995) whereas Markarian 501 showed the strongest flux 
ever detected from a blazar except for  a high state from 3C273  (McNaron-Brown et al. 1997).
These data suggest that there is a correlation between hard X-ray emission and TeV
brightness.  The hard X-ray brightness is attributed to synchrotron emission up to 
at least 100~keV.  Therefore, in Markarian 501 the synchrotron component extends 
 about a factor of 100 higher in energy (100 keV) than in Markarian 421 (1 keV), 
an unusual property  even for an X-ray selected BL Lac.

2) The EGRET detector provides only an upper limit and this implies that the 
maximum energy  output of Markarian 501 peaks in the TeV and not in the GeV regime
(see multiwavelength spectrum in Figure 9).
It is interesting to note that observations of Markarian 501 taken in 1996 by 
Kataoka et al. (1999)  show a synchrotron peak at $\rm \sim $~2~keV.  This implies 
that the synchrotron peak  of Markarian 501 shifts to higher 
energies when flaring (Figure 9).

The fact that Markarian 501 was discovered as a $\gamma$-ray source by a TeV telescope 
(Quinn et al. 1996)
(and was not an EGRET source at that time) suggests, that  besides blazars peaking 
in radio and at EGRET energies, also referred  to as radio selected BL Lacs,   the existence 
of a class of extreme blazars,  radiating most powerfully between 50~GeV to TeV 
energies (Ghisellini 1999).   
Also Markarian 421 supports this theme since it is one of the weakest nearby EGRET blazars, yet
it is detected at TeV energies.
Ghisellini (1999) suggests that even a class of  more extreme BL Lacs can exist, with the 
synchrotron emission peak at MeV energies and the inverse Compton peaking at multi- TeV energies.

\subsubsection{\underline{VHE $\gamma$-ray Spectra:} }

As mentioned in the previous section,    the understanding of the emission process
 at work requires the measurement of the  energy spectrum over a wide range of energies.  
Ideally one would like to accomplish complete coverage between X-rays to VHE $\gamma$-rays.   
Furthermore, due to the  extreme variability of blazars at all wavelengths it is crucial
to derive  accurate\footnote{The  accuracy of spectral indices should be
better than  $\pm 0.1$ to distinguish emission models.}  contemporaneous X-ray - VHE 
spectra with  dense temporal coverage.
In this section I want to emphasize the progress that has been made with existing 
Cherenkov telescopes to achieve this difficult task.  

Energy spectra for Markarian 421 and Markarian 501 have been derived by various
Cherenkov telescope groups.  Strong flaring activity as observed from Markarian 
421 in May 1996 (Gaidos et al. 1996) has resulted in the first statistically 
accurate energy spectrum between 560~GeV - 5~TeV  (Zweerink et al. 1997) showing 
no evidence for a  cutoff (also supported by large zenith angle observations 
(Krennrich et al. 1997)).  
The averaged spectrum of Markarian 501 during the flaring activity in 1997
is shown in Figure 10 (Samuelson et al. 1998) as measured by the Whipple
collaboration.   The spectrum between 260~GeV - 12~TeV is clearly curved and can 
be well fitted with a  parabolic spectrum:

\bigskip

$ \rm J(E) \propto  \: E^{-2.22
\: \pm \: 0.04_{stat.} \pm \: 0.05_{syst.} \: -(0.47 \pm 0.07)log_{10}(E) } 
\: photons \: m^{-2}  s^{-1}  TeV^{-1} $.

\bigskip

\begin{figure}
\plotone{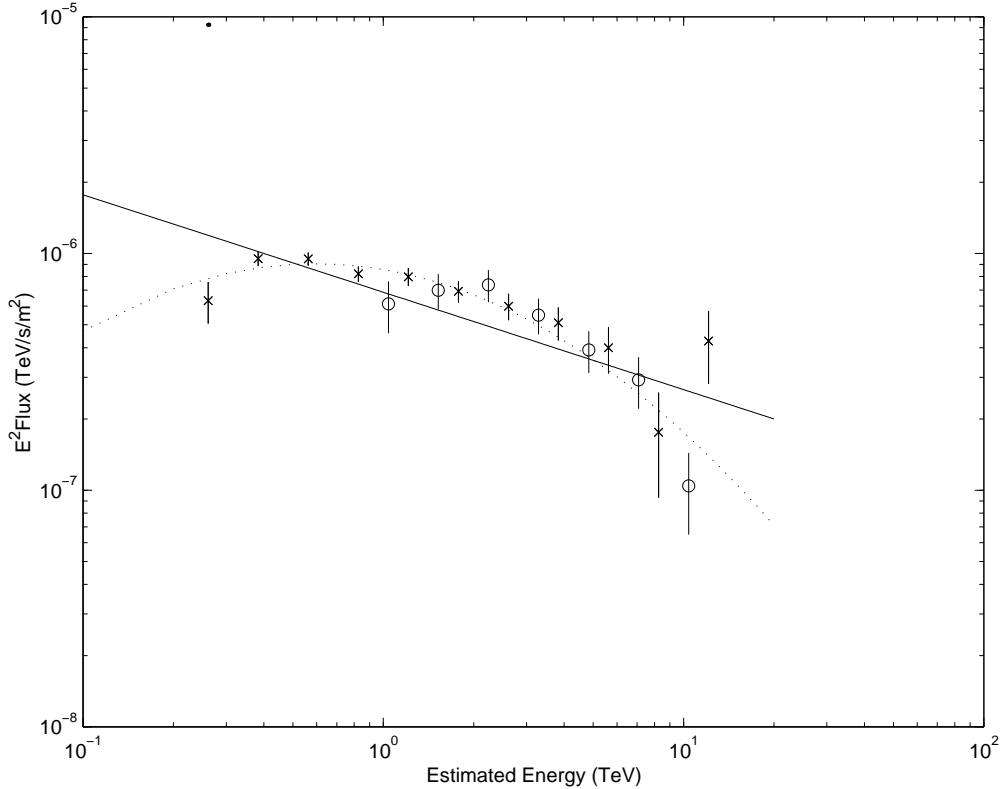}
\caption{The energy spectrum of Markarian 501 between 260~GeV - 12~TeV, measured with the
Whipple 10~m $\gamma$ray telescope (Samuelson et al. 1998).  The stars show the differential
fluxes derived from 15 hours of observations at small zenith angles, the circles show the 
fluxes derived from large zenith angle observations. 
  }
\end{figure}

\begin{figure}
\plotone{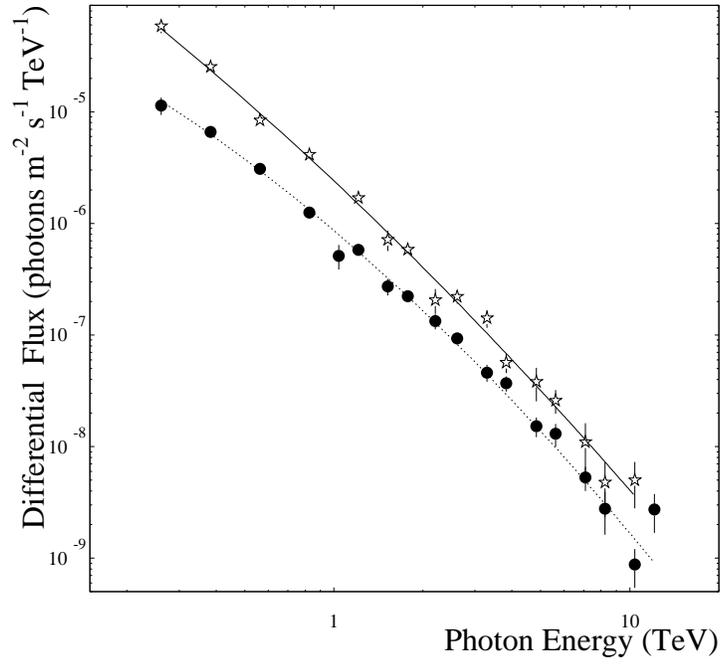}
\caption{The energy spectrum of Markarian 501 (filled circles) between 260~GeV - 12~TeV  compared  with
the spectrum of  Markarian 421 (stars) (Krennrich et al. 1999).   
  }
\end{figure}

The energy spectrum of Markarian 501 has been derived by several groups confirming
the curvature in the spectrum  (Aharonian et al. 1999b;  Djannati-Atai et al. 1999; 
Aharonian et al. 2000b) and showing good agreement between the various groups. 
Furthermore,    Aharonian et al. (1999b, 2000b) were able to extend the spectrum
in energy up to 24~TeV.  They conclude that  a cutoff in the source spectrum below 16.5 TeV 
can be excluded (Aharonian et al. 2000b).  This result by itself has major implications 
for the extragalactic IR background density.  Limits to the extragalactic
 IR density have been derived by several authors (Stanev  \& Franceschini 1998;
Biller et al. 1998;  Vassiliev 2000; Primack et al. 1999), showing that at some wavelengths
the VHE limits are a factor of 50 below limits from direct IR measurements. 
More extensive discussions of the subject  can be found elsewhere (Stanev  \& Franceschini 1998;
Biller et al. 1998;  Vassiliev 2000; Primack, Bullock, Sommerville \& MacMinn 1999; 
Catanese \& Weekes 2000; Stecker 2000; Protheroe \& Mayer 2000).

Energy spectra of comparable statistical accuracy have been derived for Markarian 421 
(Krennrich et al. 1999a) by combining data sets from various  strong flares (1-10 Crab)
in 1995 and 1996.   Figure 11 
shows the energy spectrum of Markarian 421 and Markarian 501 together in one plot.  The 
resulting spectrum for Markarian 421 is well fitted by a simple power law:

\smallskip

 $ \rm J(E) \propto  \: \:
E^{-2.54 \: \pm \: 0.03_{stat.} \: \pm \: 0.10_{syst.}} \: photons \: m^{-2} \: s^{-1}
\: TeV^{-1} $

\smallskip

This is different from the Markarian 501 spectrum, which cannot be fitted by simple 
power law, requiring a curved fit.  The spectra of Markarian 501 and Markarian 421
 are different and because the two blazars have almost the same redshift, the differences 
in their spectra must be intrinsic to the sources (or absorption near the source) and 
not due to interaction with the extragalactic  IR  background. 
The observed difference could be related to the fact that for Markarian 501, 
because the synchrotron and inverse Compton peaks are shifted to higher energies
in comparison to  Markarian 421, the detected $\gamma$-rays are closer to the peak.
At the peak one would naturally expect more curvature than further beyond the peak 
(Krennrich et al. 1999a; Djannati-Atai et al. 1999).   

Observation of Markarian 421 by  Aharonian et al. (1999c) in 1997 and 1998
 reveals a different spectrum,  showing a differential spectral index of 
$\rm 3.09 \pm 0.07 \pm 0.10$, but again
with no evidence for curvature.   
It should be emphasized that the flux levels during these observations (1997, 1998)
were only $\sim$~0.5 Crab, substantially below the flux levels of the Whipple observations
in 1995 - 1996 (1-10 Crab).    This might indicate that the energy spectrum  of Markarian 421 
becomes harder with increasing flux during flaring activity.   Further observations
are necessary to resolve this question.

Measurements of spectral variability hold great promise to deliver an additional constraint for
emission models and is therefore an important goal for  VHE observations with future
instruments.  
Evidence for variability in the spectrum of Markarian 501 has been reported by 
Piron et al.  (1999), indicating spectral softening during a period of lower emission
in 1998.   This result supports the picture that also the inverse Compton peak shifts to
lower energies as the flux decreases. This concurs well with the notion by Kataoka et al. (1999)
showing that the synchrotron peak (X-rays) was at much lower energy ($\sim $2 keV) 
during a period of low emission. 

Energy spectra are key to the understanding of the emission processes at work 
in all $\gamma$-ray sources and will continue to take an important role for
strong detections with future generation instruments.

\subsubsection{1ES2344+514, 1ES1959+650, 1ES2155-304, 3C66: }

The possibility of several more blazars detected at TeV energies is an intriguing one.
The most severe limits to IR background models could be derived if  the observation of
3C66 (Neshpor et al. 1998) were  confirmed.    For details on the other sources
see Catanese \& Weekes (2000) and references therein.
All of these sources need confirmation.

\section{Next Generation Detectors}

Existing  atmospheric Cherenkov detectors (see: Ong 1998; Catanese \& Weekes 2000)
come in a variety of designs.    The instruments fall into two categories: 
imaging and wavefront sampling  detectors.  Imaging telescopes (Weekes et al. 1989)  
record  an image of the Cherenkov  light of an air shower and use  
the image's shape, orientation  and angular position  to derive the arrival direction, primary 
energy and  type of primary particle ($\gamma$-ray, cosmic-ray nucleon or muon).
Wavefront sampling with  heliostats relies on the  Cherenkov light detection 
at various locations of the Cherenkov light pool using the light intensity and relative 
arrival times to recognize the arrival direction, primary energy and type of 
primary particle (Ong et al. 1996 ; Pare et al. 1996).   Imaging telescopes 
have an excellent sensitivity, wavefront sampling  detectors could potentially 
reach the lowest energy threshold of any ground-based instrument.

The success of the present generation imaging telescopes (CANGAROO, CAT, HEGRA, Whipple) 
has established a good sensitivity of the atmospheric imaging technique, and  it has
also become apparent that the technique can be substantially improved and the 
energy range extended to lower energies.   
A third generation of imaging detectors is currently under construction.  
The MAGIC telescope project consists of a single 17~m optical reflector
which is likely to reach an energy threshold of 30~GeV using standard
photomultipliers and 15 GeV if equipped with photodetectors containing GaAsP photocathodes
(Barrio et al. 1998). 
The other projects, CANGAROO III (Mori et al. 1999), HESS (Hofmann et al. 1999) 
and VERITAS (Weekes et al. 1999; Krennrich et al. 1999b) are based on stereoscopic imaging.
The idea of the stereo concept with multiple instruments has been first demonstrated
by Grindlay (1972).  Its realization with modern state-to-the-art 
telescopes has been achieved  by Daum et al. (1997) using the HEGRA telescope
array (4 telescope of $\rm 8.5 \: m^2$ mirror area each) showing excellent  angular resolution
and background rejection at 1~TeV (Konopelko et al. 1999).

Detailed descriptions of the various instruments are given elsewhere (e.g., see ``GeV-TeV Gamma Ray 
Astrophysics Workshop'' (1999)).  Their major properties and their impact on the various
science topics  are summarized (++ for very important; + important; o for less important) in table 2.
The most relevant improvements of the next generation of telescopes can be briefly 
summarized as follows:

\begin{itemize}
\itemsep=0pt

	\item Better flux sensitivity: the arrays of IACTs such as CANGAROO,  HESS \& VERITAS 
         are expected to
	reach flux sensitivities 10 - 20 times better than any previous installation (Weekes et al. 1999).

       \item  Reduced energy threshold: most new instruments will have energy thresholds significantly
            below 100~GeV, MAGIC will have the lowest  threshold of 30~GeV.

     \item  Energy resolution:  for spectroscopic measurements the resolution will be substantially
          improved through stereoscopic observations and improved calibration techniques.  An energy 
         resolution of  up to 10-15\% can be achieved over two decades in energy.

       \item  Angular resolution: a resolution of better than $\rm 0.1^{\deg}$ at E $>$ 100 GeV  
               ($\rm 0.03^{\deg}$ at 1 TeV)  for individual photons can be reached.  A source 
                 location capability  of $\rm 0.005^{\deg}$   will be possible for strong sources.  

       \item Larger effective area of $\rm > 0.1 \:  km^{2}$  is possible to allow measurements of 
            short flux variability.

\end{itemize}

\begin{table}
\caption{Instrument requirements for future VHE telescopes}
\begin{tabular}{lcccccc} \hline\hline
           & monitoring      &  low    & high    &  field   & energy  & angular  \\
           & capability      &  energy  & energy  &  of view &  $\rm \Delta E/E$ & resol. $\rm \sigma_{\Theta}$   \\
           & [\#of  objects] &  [GeV]     &  [TeV]    &  [$\rm {\deg}$]  &  @200 GeV &  @100 GeV \\ \hline
\underline{Instruments:}    &                 &           &           &            &                 \\ 
CANGAROO III  & 1 - 4           &    100   & $>$10  &  3 - 8   & NA  & NA      \\ 
MAGIC      & 1               &    30   & $>$10  &  3       &  20\%  &  $\rm  0.2^{\deg}$       \\ 
HESS       & 1 - 4           &    40   & $>$10  &  3 - 8  &  15\%  &  $\rm  0.1^{\deg}$     \\ 
VERITAS    & 1 - 7           &    50   & $>$10  &  3 - 10  &  15\%  & $\rm   0.09^{\deg}$   \\ \hline
\underline{Science:}    &                 &           &           &            &                 \\ 
Plerions   &   o             &  +      &   ++      &   o    &   ++    &  +            \\
Pulsars    &   o             &  ++     &   o      &   o    &  ++   & o               \\
Shell-type SNRs       &   o             &  +      &   ++     &   ++   &  ++  & ++     \\
EGRET                 &                 &         &          &               \\
Unidentified         &   o             &  ++     &   +      &   ++   &  ++  & ++          \\
Gal. Diffuse   &   o             &  ++     &   +      &   ++   &  ++     & +      \\
Gal. Plane Survey     &   o             &  ++      &   o      &   ++   &  + & ++         \\
Blazars $z<0.1$ &   ++            &  +      &   ++     &   o   &  ++   & o     \\
Blazars $z > 0.1$  &   ++            &  ++     &   +      &   o    &  ++  & o     \\
Gamma Ray  &                 &         &          &               \\
Bursts     &   ++            &  ++     &   +      &   o    &  ++  & +     \\
Extragalactic IR         &                 &         &          &               \\
Background &   ++            &  ++     &   ++     &   o    &  ++   & o    \\
Supersymmetric  &                 &         &          &               \\
particle decay  &  o            &  ++     &  o        &   +    & ++  & +           \\

\end{tabular}
\end{table}

The scientific potential for the various science topics can be summarized as follows:

\underline{Plerions:} A VERITAS-like detector with a flux sensitivity of 0.5\% of 
         the Crab should be capable of  detecting Crab-like plerions out to a distance 
         of 20~kpc (see Weekes et al. 1999).   This will  allow a search for synchrotron 
         self-Compton dominated plerions (Crab-like) over 2/3 of our galaxy.

\underline{Pulsars:}    Because of the potentially sharp cutoff in pulsar energy spectra 
                        an energy threshold of $<$ 50~GeV 
                        is desirable.   The excellent sensitivity due to a large collection area
                         and excellent energy resolution  of the next generation imaging telescopes
                         holds great promise in resolving the question as to where $\gamma$-rays 
                         in pulsars originate.

\underline{Shell-type SNRs:} Next generation telescope arrays with their  arc-minute angular resolution
                             will be capable of mapping the  sites of putative cosmic-ray acceleration 
                            and correlate   them with interstellar matter density  and compare them 
                            with their X-ray luminosity.   
                            The versatility of telescope arrays allows a wide-field-of-view mode 
                            for  the study of more extended SNRs. 
                            The good energy resolution of arrays will help to measure the energy spectrum
                            from 50~GeV to 10  TeV.   Ultimately a combined spectrum including data from 
                            GLAST  and VHE telescopes will be a powerful tool to separate $\gamma$-ray
                            emission of IC origin from nucleonic origin.

\underline{Galactic diffuse emission:}  The intrinsically  large collection area of ground-based telescope
                                        combined with a low energy threshold and a good energy resolution
                                        should allow the measurement of the energy spectrum
                                        of the galactic diffuse emission from different regions
                                        in the galaxy  in the range of 50 GeV - TeV.    
                                          Arrays of Cherenkov telescopes increase the  
                                         field-of-view by the use of sub-arrays
                                        (e.g., VERITAS: two arrays of 3 telescopes).

\underline{Blazars:}  Ground-based Cherenkov telescopes have the highest sensitivity to short
                       $\gamma$-ray flares from blazars because of their large collection  areas.
                      Thus, detailed multiwavelength studies on scales of minutes to hours 
                      is the domain of the X-ray and ground-based VHE instruments. 
                      Maybe most important will be multiwavelength studies that include the 
                      whole $\gamma$-ray spectrum from MeV energies to ten's of TeV and
                      a dense sampling  involving GLAST together with a number of Cherenkov 
                       telescopes at different geographical longitude.
                      It is important to realize that observations with  the next generation 
                      space telescope  GLAST covering 20~MeV - 300~GeV  together with ground-based 
                      telescopes (e.g., VERITAS: 50~GeV-50~TeV) will provide for the first time
                       a significant overlap in energy.  A cross-calibration between the 
                      ground-based and satellite telescopes will allow the derivation of
                      detailed energy spectra over 6 orders of magnitude in energy.

\section{Acknowledgement}
I would like to thank the organizers of this excellent conference and their hospitality.
Also many thanks to Stephane Le Bohec, David Carter-Lewis and Trevor Weekes
 for reading the manuscript and Olaf  Reimer for help with a figure.
This research is supported by grants from the U.S. Department of Energy.


\newpage


\begin{references}

\reference 

Aharonian, F.A. et al. (1999a), A\&A, 346, 913

Aharonian, F.A. et al. (1999b), A\&A, in press (astro-ph/99030455)

Aharonian, F.A. et al. (1999c), A\&A, in press (astro-ph/9905032)

Aharonian, F.A. et al.  (2000a), ApJ, 539, 317

Aharonian, F.A. et al. (2000b), A\&A, in press (astro-ph/0011483)

Aharonian, F.A., \& Atoyan, A.M. (1999), A\&A, in press

Aharonian, F.A., (2000), New Astronomy, 2000, vol.5, pp. 377-395

Akerlof, C.W. et al. (1993), A\&A, 274, L17

Allen, G.E., Gotthelf, E.V., Petre, R. (1999), Proc. of the 26th ICRC (Salt Lake City), 3, 480

Amelino-Camelia, G. et al. (1998), Nature, 383, 319

Atoyan, A.M., Aharonian, F.A. (1996), MMRAS, 278, 525

Baring, M.G. (1999), Proc. of GeV - TeV Gamma Ray Astrophysics Workshop (Snowbird),
                   eds B. Dingus et al.,  173

Barrio, J.A. et al. (1998), ``The MAGIC telescope'', MPI-Ph/E98-5

Becker, W. et al. (1994), A\&A, in press (astro-ph/9501007)

Bednarek, W., \& Protheroe, R.J. (1997), Phys. Rev. Lett., 79, 2616

Bell, A.R., (1978), MNRAS, 182, 147

Bergstr\"om, L., Ullio, P., Buckley, J.H. (1998), Astrop. Phys., 9, 137

Bertsch, D.L. et al. (1993), ApJ, 416, 587

Biermann, P.L., \& Strom, R. (1993), A\&A, 275, 659

Biller, S.D.  et al. (1998), Phys. Rev. Lett., 80, 2992

Biller, S.D. et al. (1999), Phys. Rev. Lett., 83, 2108

Blandford, R.D., \& Rees, M.J. (1978), in Pittsburgh Conf. on BL LAC Objects, ed. A.M. Wolfe 
                  (Pittsburgh: Univ. Pittsburgh Press), 328

Blandford, R.D., \& Ostriker, J.P. (1978), ApJ, 221, L29

Blandford, R.D., \& Levinson, A. (1995), ApJ, 441, 79

B\"ottcher, M. \& Dermer, C.D. (1998), ApJ, 499, L131

Buckley, J.H. et al. (1996), ApJ, 472, L9

Buckley, J.H. et al. (1998), A\&A, 329, 639

Catanese, M.A. et al. (1997), ApJ, 487, L143

Catanese, M.A. \& Weekes, T.C. (2000), PASP, in press

Chadwick, P.M. (1997),  in Proc. of the 25th Internat. Cosmic Ray Conf. (Durban, South Africa),
                   eds. M.S. Potgieter, B.C. Raubenheimer, D.J. van der Walt, 3, 189

Chadwick, P.M. (1999),  Proc. of GeV - TeV Gamma Ray Astrophysics Workshop (Snowbird),
                   eds B. Dingus et al., 210

Cheng, K.S., Ho, C., \& Ruderman, M.A. (1984), ApJ, 300, 500

Daugherty, J.K., \& Harding, A.K. (1996), ApJ, 458, 278

Djannati-Atai, A. et al. (1999), A\&A, 350, 17

Drury, L.O'C, Aharonian, F.A., V\"olk, H.J. (1994), A\&A, 287, 959 

De Jager, O.C., \& Harding, A.K. (1992), ApJ, 396, 161 

De Jager, O.C., et al. (1996), ApJ, 457, 253

De Jager, O.C. et al. (2000),  Proc. of the Int. Symposium on 
                               High Energy Gamma-Ray Astronomy (Heidelberg),
                               eds. Aharonian, F.A., V\"olk, H., in press


Daum, A. et al. (1997), Astrop. Physics, 8, 1

De Naurois, M. et al. (2000),  Proc. of the Int. Symposium 
                             on High Energy Gamma-Ray Astronomy (Heidelberg),
                              eds. Aharonian, F.A., V\"olk, H., in press


Derdeyn, S.M. et al. (1972), Nucl. Instr. and Meth., 98, 557

Dermer, C.D., Schlickeiser, R. \& Mastichiadis, A. (1992), A\&A, 256, L27

Esposito, J.A. et al.  (1996), ApJ, 461, 820

Fichtel, C.E. et al. (1975), ApJ, 198, 163

Gaidos, J.A. et al. (1996), Nature, 383, 319

Gaisser, T.K., Protheroe, R.J. \& Stanev, T.  (1998), ApJ, 492, 219

Ghisellini, G. (1999), Astrop. Physics, 11, 11

Grenier, I.A. (1999),   Proc. of GeV - TeV Gamma Ray Astrophysics Workshop (Snowbird),
                   eds B. Dingus et al., 261

Grindlay, J. (1972), ApJ, 174, L9

Harding, A.K. (1996), Space Science Reviews, 75, 257

Hartmann, R.C., et al. (1999), ApJS, 123, 79


Hillas, A.M. et al. (1998), ApJ, 503, 744

Hofmann, W. et al. (1999),  Proc. of GeV - TeV Gamma Ray Astrophysics Workshop (Snowbird),
                   eds B. Dingus et al., 500

Hunter, S.D., et al. (1997), ApJ, 481, 205

Jaffe, T.R. (1997), ApJ, 484, L129

Kataoka, J. et al. (1999), ApJ, 514, 138

Kifune, T. et al. (1995), ApJ, 438, L91

Konopelko, A. et al. (1999), Astroparticle Physics, 11, 135

Koyama, M. et al.  (1995), Nature, 378, 255

Kraushaar, W.L.  et al. (1972), ApJ, 177, 341

Krennrich,  F. et al. (1997), 
             in Proc. of Workshop on TeV-Astrophysics, Kruger National Park, 77 

Krennrich,  F. et al. (1997), ApJ, 481, 758

Krennrich,  F. et al. (1999a), ApJ, 511, 149

Krennrich,  F. et al. (1999b),  Proceedings of GeV - TeV Gamma Ray Astrophysics Workshop (Snowbird),
                   eds B. Dingus et al., 515

Krennrich, F., Le Bohec, S.,  \& Weekes, T.C. (2000), ApJ, 529, 506 

Lamb, R.C., \& Macomb, D.J. (1997), ApJ, 488, 872

Le Bohec, S. et al. (2000), ApJ, 539, 209

Legage, P.O., \& Cesarski, C.J. (1983), A\&A, 118, 223

Lessard, R.W. et al. (2000), ApJ, 531, 942

Macomb, D.J. et al. (1995), ApJ, 449, L99

Macomb, D.J. et al. (1996), ApJ, 459, L111

Mannheim, K. \& Biermann, P.L. (1992), A\&A, 253, L21

Mannheim, K. (1993), A\&A, 269, 67

Maraschi, L., et al. (1999), ApJ, 526, L81

Maraschi, L., Ghisellini, G. \& Celotti, A. (1992), ApJ, 397, L5

Marscher, A.P., \& Bloom, S.D. (1994), in the Second Compton Symposium, eds.
                      C.E. Fichtel, N. Gehrels \& J.P. Norris (New York: AIP), 573



Mastichiadis, A. (1996), A\&A, 305, L53

Mastichiadis, A. \& de Jager, O.C. (1996),  A\&A, 311, L5

Mayer-Hasselwander, H.A. et al. (1982), A\&A, 105, 164

McNaron-Brown, K., et al. (1995), ApJ, 451, 575

McNaron-Brown, K., et al. (1997), ApJ, 474, L85

Michelson, P.F. et al. (1999), GLAST, Large Area Telescope Flight Investigation,
                    (http://www-glast.stanford.edu/pubfiles/proposals/bigprop/

Muraishi, H. et al. (2000), A\&A, in press

Mori, M., (1997), ApJ, 478, 225

Mori, M. et al. (1999),  Proceedings of GeV - TeV Gamma Ray Astrophysics Workshop (Snowbird),
                   eds B. Dingus et al., 485

Moskalenko, I.V., Strong, A.W. \& Reimer, O. (1998), A\&A, 338, L75

M\"ucke, A., \& Protheroe, R.J. (2000), Astrop. Physics, in press 

Naito, T. \& Takahara, F. (1994), J.Phys. G, Nucl. Part. Phys., 20, 477

Neshpor, Y.I. et al. (1998), Astron. Lett., 24, 134

Nolan, P.L. et al. (1993), ApJ, 409, 697

Ong, R.A. et al. (1996), Nuclear Instruments and Methods, Volume A408, 468

Ong, R.A. (1998), Phys. Rep., 305, 93

Ormes, J.F., Digel, S., Moskalenko, I.V. \& Williamson, R. (2000), AIP Conf.Proc., 528, 445-448

Pare, E. et al. (1996), Space Science Reviews, 75, 127

Pfeffermann, E., \& Aschenbach, B. (1996), in R\"ontgenstrahlung from the Universe, Int.
                    Conf. on X-ray Astronomy and Astrophysics, eds. H.U. Zimmermann,
                    J.E. Tr\"umper, H. Yorke, MPE Report 263, P267

Pian, E. et al. (1998), ApJ, 492, L17

Piron, F. et al. (1999),   Proc. of GeV - TeV Gamma Ray Astrophysics Workshop (Snowbird),
                   eds B. Dingus et al., 113

Pohl, M. et al. (1996), A\&A, 307, 57

Pohl, M. et al. (1997), ApJ, 491, 159

Pohl, M. \& Esposito, J.A. (1998), ApJ, 507, 327 

Primack, J.R., Bullock, J.S., Sommerville, R.S. \& MacMinn, D. (1999),
                            Astrop.  Physics, 11,  93
            

Protheroe, R. J., Bhat, C.L., Fleury, P. \& Lorenz, E. (1997), in Proc. of the 25th 
                   Internat. Cosmic Ray Conf. (Durban, South Africa),
                   eds. M.S. Potgieter, B.C. Raubenheimer, D.J. van der Walt, 8, 317

Protheroe, R. J. \& Meyer, H. (2000),  Phys. Lett., B493,  1-6

P\"uhlhofer, G. et al. (1999), Proc.  of the 26th ICRC (Salt Lake City), 3, 492

Quinn, J. et al. (1996), ApJ, 456, L83

Quinn, J. et al. (1999), ApJ, 518, 693


Romani, R.W. (1996), ApJ, 470, 469

Samuelson, F.W. (1998), ApJ, 501, L17

Sako, T.  et al. (2000), ApJ, 537, 422

Scarsi, L. et al. (1977), Proc. 12th ESLAB Symp. ESA-SP, 124, 3

Sikora, M., Begelman, M.C., \& Rees, M.J. (1994), ApJ, 421, 153

Srinivasan, R. et al. (1997), ApJ, 489, 170

Stanev, T. \& Franceschini, A. (1998), ApJ, 494, L159

Stecker, F.W. (2000), Proc. IAU Symp. 204 "The Extragalactic Background and its
     Cosmological Implications", eds. M. Harwit and M.G. Hauser, in press

Sturner, S.J., \& Dermer, C.D. (1995), A\&A, 293, L17

Strong, A.W., Moskalenko, I.V., \& Reimer, O.  (2000), ApJ, 537, 763

Swordy, S.P. et al. (1990), ApJ, 349, 625


Tanimori, T. et al. (1998a), ApJ, 492, L33

Tanimori, T. et al. (1998b), ApJ, 497, L25

Tanimori, T. et al. (1999), ApJ, 492, L33

Takahashi, T. et al. (1999), Astroparticle Physics, 11, 177

Thompson, D.J. (1997), in Neutron Stars and Pulsars, eds. Shibazaki, N., et al.
                         (Tokyo: Univ. Acad. Press), 273

Vassiliev, V.V. (2000), Astrop. Physics, in press

Weekes, T.C. et al. (1989), ApJ, 342, 379

Weekes, T.C. (1999),  Proceedings of GeV - TeV Gamma Ray Astrophysics Workshop (Snowbird),
                   eds B. Dingus et al., 3

Weekes, T.C. (2000), Proc, of the International Symposium on High Energy Gamma-Ray Astronomy (Heidelberg),
                  eds. Aharonian, F.A., V\"olk, H., in press

Yoshida, T., \& Yanagita, S. (1997), in Proc. Second INTEGRAL workshop on the Transparent Universe (SP-382; 
               Paris; ESA), 85

Yoshikoshi, T. et al.  (1997), ApJ, 487, L65

Zweerink, J.A. et al. (1997), ApJ, 490, L141

\end{references}
\end{document}